\documentclass[aps,prb,twocolumn,showpacs]{revtex4}
\newcommand{\bsim}{\mbox{\raisebox{-0.1cm}{$\;
\stackrel{\textstyle>}{\sim}\;$}}}
\newcommand{\lsim}{\mbox{\raisebox{-0.1cm}{$\;
\stackrel{\textstyle<}{\sim}\;$}}}
\usepackage{psfig}
\usepackage{graphicx}

\begin{document}

\title{Isotope effects in the Hubbard-Holstein model within
dynamical mean-field theory}

\author{P. Paci,$^1$ M. Capone,$^{2,3}$ E. Cappelluti,$^{2,3}$
S. Ciuchi,$^{4,2}$ and C. Grimaldi$^{5,6}$}

\affiliation{$^1$International School for Advanced Studies
(SISSA), via Beirut 2-4, 34014 Trieste, Italy}

\affiliation{$^2$SMC-INFM and Istituto dei Sistemi Complessi, CNR,
via dei Taurini 19, 00185 Roma, Italy}

\affiliation{$^3$Dipart. di Fisica, Universit\`a ``La Sapienza'',
P.le A. Moro 2, 00185 Roma, Italy}

\affiliation{$^4$Dipart. di Fisica, Universit\`{a} de L'Aquila,
and INFM UdR AQ, 67010 Coppito-L'Aquila, Italy}

\affiliation{$^5$Ecole Polytechnique F\'ed\'erale de Lausanne,
LPM, Station 17, CH-1015 Lausanne, Switzerland}

\affiliation{$^6$DPMC, Universit\'e de Gen\`eve, 24 quai Ernest
Ansermet, CH-1211 Gen\`eve 4, Switzerland}
\date{\today}

\begin{abstract}
We study the isotope effects arising from the coupling of correlated electrons
with dispersionless phonons by considering the Hubbard-Holstein model 
at half-filling within the dynamical mean-field theory. 
In particular we calculate the
isotope effects on the quasi-particle spectral weight $Z$, the renormalized
phonon frequency, and the static charge and spin
susceptibilities. In the weakly correlated regime $U/t \lesssim 1.5$, where $U$
is the Hubbard repulsion and $t$ is the bare electron half-bandwidth, the
physical properties are qualitatively similar to those characterizing the 
Holstein model in the absence of Coulomb repulsion, where the bipolaronic
binding takes place at large electron-phonon coupling, and it reflects in
divergent isotope responses. On the contrary in the strongly correlated regime
$U/t \gtrsim 1.5$, where the bipolaronic metal-insulator transition becomes of 
first order, the isotope effects are bounded, suggesting that the first order
transition is  likely driven by an electronic mechanism, rather then by a
lattice instability. These results point out how the isotope responses are
extremely sensitive to phase boundaries and they may be used to characterize
the competition between the electron-phonon coupling and the Hubbard repulsion.
\end{abstract}
\maketitle

\section{Introduction}

One of the most direct methods to establish the coupling of an
electronic property with the dynamics of the underlying lattice
is the isotope effect (IE), $\alpha_A = - d\ln A / d\ln M$, which
probes the dependence of the electronic quantity $A$ on the ion
mass $M$. For example, the observation of  a large IE
on the superconducting transition temperature $T_c$ has been an
important evidence for a phonon-mediated pairing mechanism in
conventional low-temperature superconductors. In the
weak-coupling BCS model, the superconducting critical temperature
is proportional to the Debye phonon frequency which scales as
$M^{-1/2}$.
The resulting IE on $T_c$,
$\alpha_{T_c}=1/2$, has been considered as one of the most
important confirmation of the BCS theory.

The vanishingly small values of $\alpha_{T_c}$ in optimally doped
high-$T_c$ copper oxides superconductors has initially induced  a
large part of the scientific community to believe that the
pairing mechanism in these compounds was mainly of electronic
origin, and oriented the theoretical research towards purely
electronic models, relegating the lattice to a secondary role.
Later experimental investigations showed however that small
values of $\alpha_{T_c}$ are peculiar of the optimal doping,
whereas sizable values of $\alpha_{T_c}$, even larger than the BCS
prediction, were reported in the underdoped phase.\cite{franck}
Furthermore, the improvement of the experimental accuracy has
allowed recently to establish the presence of isotopic shifts
also in several physical quantities different from the
superconducting $T_c$.

Sizable isotope effects (IEs) have been indeed found on the zero
temperature penetration depth $\lambda_{\rm
L}(0)$,\cite{keller1,keller2} on the Knight shift,\cite{mali1} on
the nuclear quadrupole resonance,\cite{mali2} on the pseudogap
temperature,\cite{rubiotemprano} and on the
 angular resolved photoemission spectra.\cite{lanzara} 
These findings are
particularly remarkable since the Migdal-Eliashberg theory of
the electron-phonon interaction predicts vanishingly small
IEs other than $\alpha_{T_c}$. For example, according
to the BCS description of the superconducting state, the zero
temperature penetration depth is given by $\lambda_{\rm
L}(0)=(m^*/n_s)^{1/2}$, where $n_s$ is the density of the
condensate and $m^*$ is the effective electron mass. According to
the Migdal-Eliashberg theory, $m^*=(1+\lambda)m$, where $m$ is the bare band
mass and $\lambda$ is the electron-phonon coupling constant.
Since $\lambda$ is independent of the ion mass, the isotope
effect on the penetration depth is expected to be zero, in
contrast with Refs. \onlinecite{keller1,keller2}.

The completely anomalous set of isotope dependences signals
a relevant electron-phonon coupling, but at the same time highlights
that the interplay with strong correlations and possibly with other
features of the cuprates needs to be taken into account.
Different theoretical models have been proposed to account for
these anomalous IEs.\cite{gcp,chi,millis,schneider,korni,bussmann,grilli}
The analysis has been mainly concentrated
on purely electron-phonon narrow-band systems, as the Holstein model which
however neglects possible effect of the anisotropy of the electron-phonon 
interaction considered in  Ref. \onlinecite{Devereaux}.

The IEs on $m^*$ in interacting electron-phonon systems 
in the absence of electron-electron interactions have been
investigated in details in Refs. \onlinecite{gcp,chi,millis}.
More recently, the ability of
the IEs in revealing different physical regimes of electron-phonon
systems has been discussed in Ref. \onlinecite{paci} where remarkable anomalous
IEs on $m^*$ and on the dressed phonon frequency $\Omega_0$ were
reported as the system enters in a polaronic regime. In order to focus on the
metallic properties and to clarify the origin of these anomalous IEs, however,
a spinless Holstein model in the DMFT approximation
was considered there, which enforces the metallic
character in the whole space of parameters.

In the opposite case of strong correlations the electron-phonon interaction 
in the weak-intermediate limit
has been treated for instance by means of an adiabatic theory
on small clusters,\cite{gunnarsson} or of coherent potential
approximation.\cite{fratini}
In this case the IE reflects in a smooth modification of high
energy features such as the Hubbard bands. In the strongly correlated
antiferromagnetic phase IEs on the single hole spectral properties
have been also analyzed with diagrammatic
quantum Monte Carlo techniques.\cite{nagaosa}

The scope of this paper is to provide a nonperturbative analysis 
of the isotope effects on different observables in a regime in which 
strong correlation and intermediate/strong electron-phonon interaction 
coexist. 
To this aim we employ the dynamical mean-field technique (DMFT),
which is powerful tool to investigate
the nonperturbative regimes of strongly correlated and electron-phonon systems.
In particular,
in this paper we consider the metallic regime of the Hubbard-Holstein 
model, deliberately excluding broken-symmetry phases.
We also limit ourselves to the half-filled systems, 
where in DMFT the correlation-driven Mott transition 
or the electron-phonon driven pairing transition take place
as one or the
other coupling becomes large, and we follow the behavior of the isotope effects
when these phase boundaries are approached. 
It is well known that DMFT becomes exact only in the limit
of infinite coordination number, and it has to be viewed as an approximate
method
in finite dimensions. Yet, DMFT allows us to access the full local quantum
dynamics,
and to deal with many different energy scales.
Moreover the assumption of a local self-energy, 
implicit in the DMFT, makes this technique not suitable to investigate
the momentum dependence of the physical quantities.
This is for instance the case of
Fr\"olich-like electron-phonon Hamiltonians and
of the long-range Coulomb interaction, which gives rise
to important screening effects here neglected.\cite{ginzburg}
This is also the case of the momentum dependence of the
electron-phonon interaction induced by the electronic
correlation itself\cite{kulic,kulic2,ccp},
which cannot be properly take into account
within the context of the DMFT. 

A general belief is that the electronic correlation competes with the 
electron-phonon interaction leading to a reduction of the phonon effects on 
the electronic properties with a consequent decrease of the magnitude of the 
possible IEs. We show that in particular regimes, close to phase
boundaries, the Hubbard repulsion can actually
{\em enhance} the dependence of many physical quantities on the phonon
frequency, yielding an increase of the IEs. Indeed the isotope
effects turns out to be extremely sensitive to the closeness of metal-insulator
transitions, and emerge as a tool to
investigate the physics underlying the different regimes of the phase diagram.

This paper is organized as follows. In Sec. \ref{approach} we
introduce the Hubbard-Holstein model and the dynamical mean field
method. In Sec. \ref{s-weak} and Sec. \ref{s-strong} we report our
results for the weakly correlated and strongly correlated
regimes, respectively, while a study of the IEs as a
function of the Hubbard-$U$ repulsion is the subject of
Sec. \ref{U-dependence}. The conclusions are summarized in
Sec. \ref{concl}

\section{Our approach}
\label{approach}

The Hubbard-Holstein model is defined by the Hamiltonian:
\begin{eqnarray}
H&=&
-t \sum_{\langle ij\rangle \sigma}
c_{i\sigma}^\dagger c_{j\sigma}
+ U \sum_{i} n_{i\uparrow} n_{i\downarrow}
\nonumber\\
&&
+g \sum_{i\sigma} n_{i\sigma} \left(a_i+a_i^\dagger \right)
+\omega_0 \sum_i a_i^\dagger a_i,
\end{eqnarray}
where $t$ is the hopping amplitude between nearest neighboring sites, 
$U$ is the Hubbard electron-electron repulsion, $g$ is the
electron-phonon matrix element, and $\omega_0$ the Einstein
phonon frequency. 

We can identify three independent dimensionless
parameters, the electron-phonon coupling constant $\lambda=2
g^2/(\omega_0 t)$, the adiabatic ratio $\gamma=\omega_0/t$, which
parametrizes the role of the lattice quantum fluctuations, and
the dimensionless Hubbard repulsion $U/t$. It is important to
note that  $\omega_0 \propto 1/\sqrt{M}$
and $g \propto 1/\sqrt{M\omega_0}$, where $M$ is the ionic mass,
so that the only physical quantity which depends on $M$ is the
adiabatic ratio $\gamma$. Finite IEs stem thus from
an explicit dependence on $\gamma$ and they are hence a direct
signal of the role of the lattice quantum fluctuations.
The locality of the interaction terms makes the
Hubbard-Holstein model particularly
suitable to be investigated by dynamical
mean-field theory.\cite{revdmft} 
In this approach the spatial fluctuations
are frozen, i.e., all the lattice sites are assumed to be equivalent,
but the local electron and phonon dynamics are fully accounted for.
In this sense the method generalizes to a quantum framework the
classical mean-field theories, and it becomes exact in the infinite
coordination limit.
Hence the lattice model is mapped onto a quantum impurity model in
which an interacting site is embedded into a non-interacting bath,
whose spectral function has to be self-consistently determined.
In our case, the impurity site displays both Coulomb repulsion and
a coupling to a local phonon.
The self-consistency equation contains all the information about
the original lattice model through the bare density of states.
As often, here we consider an infinite coordination Bethe
lattice with half-bandwidth $t$,\cite{notat} 
which reproduces the finite bandwidth of finite-dimensional systems. 
For the normal paramagnetic phase, 
the self-consistency condition on the Bethe lattice reads
\begin{equation}
\Delta(\omega) = \frac{t^2}{4}G(\omega),
\end{equation}
where $G$ is the Green's function on the impurity site, which corresponds
to the local component of the lattice Green's function,
and $\Delta$ is the bath hybridization function. Within DMFT the electron and
phonon self-energies are local, i.e., momentum independent.

To solve the impurity model we use the exact diagonalization 
method,\cite{caffarel} in which the dynamical DMFT bath
is described in terms of a finite set $N_s$ of Anderson
impurity levels, while a finite cut-off is imposed on the
total number $N_{\rm ph}$ of the phononic Hilbert space.
Typical values we considered are $N_s = 9,10$, and $N_{\rm ph} \sim 20$.
Both these quantities were varied to check the effective
stability of the results. The self-consistent DMFT set of equations
was solved in the Matsubara space where a fictitious temperature
$\tilde{T}$ plays the role of an energy cut-off.
Particularly small value of $\tilde{T}$ up to
$\tilde{T}/t \sim 1/400$ were needed to ensure a high accuracy of
our results.

In this paper we investigate unconventional IEs on
both single-particle and two-particle quantities. In particular,
concerning the one-particle properties, we focus on
the quasi-particle spectral weight $Z$, which
in DMFT coincides with the inverse of the
effective electronic mass $m^*/m$ due to the momentum independence of 
the self-energy $\Sigma$
\begin{equation}
\label{mass} 
Z =
\left ( 1-
\left.\frac{\partial\Sigma(\omega_n)}{\partial\omega_n}\right|_{\omega_n=0}
\right)^{-1},
\end{equation}
and on the renormalized phonon frequency $\Omega_0 $ obtained as\cite{Max1}
\begin{equation}
\label{omega}
\left(\frac{\Omega_0}{\omega_0}\right)^2=-\frac{2D^{-1}(\omega_m=0)}{\omega_0}.
\end{equation}
In the above expressions, $\Sigma(\omega_n)$ is the electronic
self-energy and $D(\omega_m)$ is the phonon propagator
$D(\omega_m) = -\langle x(\omega_m) x(-\omega_m) \rangle$,
where $x=\left(a+a^\dagger\right)$ is the
dimensionless lattice coordinate operator.
$\omega_n$ and $\omega_m$
are respectively fermionic $\omega_n=\pi \tilde{T}(2n+1)$ and
bosonic $\omega_m=2\pi \tilde{T}m$ Matsubara frequencies. 
In practice $Z$ is computed by linearly extrapolating from the first
Matsubara frequency.

The knowledge of the renormalized phonon frequency $\Omega_0$ allows
us to evaluate also the static component of the {\it local} charge 
susceptibility $\rho = \rho(\omega_m = 0)$, where 
$\rho(\omega_m) = \langle
n(\omega_m) n(-\omega_m) \rangle$, $n$ being the density operator on the 
impurity.
It is indeed easy to prove that $\rho=(1-\omega_0^2/\Omega_0^2)/\lambda t$. 
We evaluate also the
local static spin susceptibility $\chi = \chi(\omega_{m}=0)$ for
the spin response function $\chi(\omega_m) = \langle
S_z(\omega_m) S_z(-\omega_m) \rangle$, where $S_z =
(1/2)\sum_{\alpha\beta}c_\alpha^\dagger
\sigma^z_{\alpha\beta}c_\beta$ is the $z$-component of the local
electron spin ($\hat{\sigma}^z$ being the Pauli matrix
$\hat{\tau}_3$). Finally, we consider also the local lattice
probability distribution function $P(x)=|\langle\phi_0|x\rangle|^2$,
where $|x\rangle$ is the eigenstate of the
lattice coordinate operator $x$.
In our exact diagonalization approach we can evaluate
$P(x)=\sum_{n,m}
\psi_n(x)\psi_m(x)\langle \phi_0 | n\rangle \langle m |\phi_0
\rangle $, where $|\phi_0\rangle$ is the ground state,
$| n\rangle$ are the eigenstates of the
harmonic oscillator and $\psi_n(x)$ the corresponding
eigenfunctions.

Despite its simplicity, the Hubbard-Holstein model has quite a rich
phase diagram, in which different phases and regimes appear.
In order to have a better comparison between the different
regimes, we organize our discussion according to degree of correlation.
In the next section we
focus on the weakly correlated regime of the Holstein-Hubbard
model, where the presence of the on-site Hubbard repulsion changes
only quantitatively the results of the pure Holstein spinful
model, but it does not give rise to new features. We then discuss
in section \ref{s-strong} the regime of strong electronic
correlation where the interplay between electron-phonon and
electron-electron interaction induces qualitative changes on the
behavior of the IEs.

\section{Weakly correlated regime}
\label{s-weak}

Already in the absence of correlation, the introduction of the spin degeneracy
introduces additional physics with respect to the spinless system,
whose isotope
effects have been studied in Ref. \onlinecite{paci}.
In the spinless case
indeed the polaron crossover is not
associated with a metal-insulator transition (MIT) which is only asymptotically
approached for $\lambda \rightarrow \infty$. The persistence of a metallic
character in the polaronic phase is reflected in the finite value
of both the quasi-particle weight and the renormalized phonon frequency.\cite{Max1} On the
other hand in the spinful case the residual attractive interaction between the
electronic or polaronic charges leads
to the possibility to form a insulating
phase of localized pairs which in the adiabatic regime ($\gamma \ll 1$)
are accompanied by strong lattice deformation
(bipolarons).\cite{carta} 
This occurs within the DMFT approximation
for $\lambda$ larger than a finite critical value where the
quasi-particle weight vanishes at $\lambda_{\rm MIT}$ signaling a
pair-MIT.\cite{note-infinte-d}
In the adiabatic limit ($\gamma=0$) the
pair-MIT has a precursor in the phonon softening which occurs at
$\lambda_{\rm ph}$.\cite{millis2} 
As the bare phonon frequency is increased, the  phonon
softening tends to occur more closely to the pair-MIT.\cite{koller}

\begin{figure}[!t!]
\centerline{\psfig{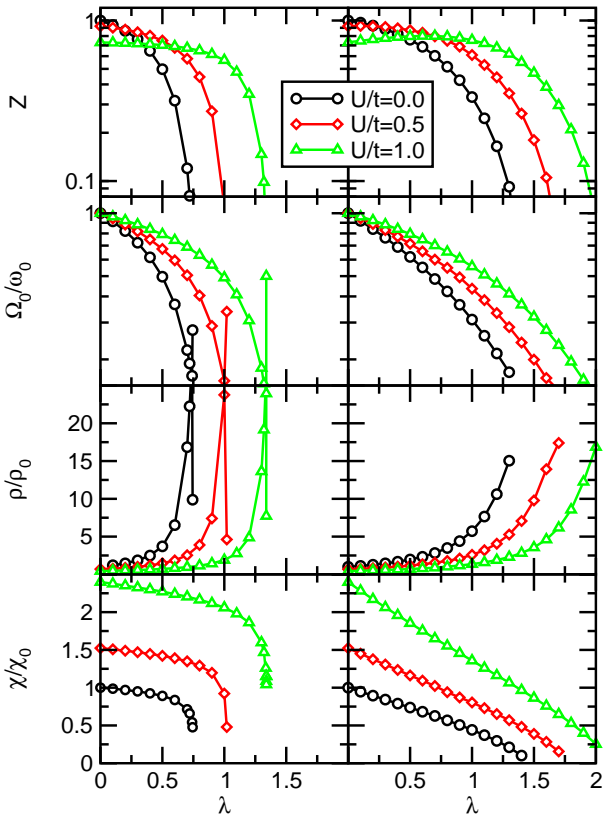}}
\caption{(color online) Behavior of the quasi-particle
spectral weight $Z$, the
renormalized phonon frequency $\Omega_0$, the local
static charge susceptibility $\rho$ and the local
static spin susceptibility $\chi$
as functions of
the electron-phonon coupling $\lambda$ for different weak-intermediate
values
of the Hubbard repulsion $U$. In the left panel
the curves are plotted for the adiabatic ratio
$\gamma=0.1$ while in the right panel the adiabatic ratio is
$\gamma=1.0$.}
\label{f-Uweak}
\end{figure}

In Fig. \ref{f-Uweak} we show the quasi-particle spectral
weight $Z$, the renormalized phonon
frequency $\Omega_0$, the local static spin susceptibility $\chi$ 
and the local static
charge susceptibility $\rho$ as functions of the electron-phonon coupling
$\lambda$ for different weak-intermediate values of the Hubbard repulsion $U$
and for adiabatic ratios $\gamma=0.1, 1.0$. $\chi_0$ and $\rho_0$ represent
respectively the local spin and charge susceptibility in the absence of
electron-phonon and Hubbard interaction. As mentioned above, the main
difference here with respect to the spinless Holstein model is the possibility
to form local bipolaronic pairs of electrons with opposite spin. This leads to
a metal-insulator transition at a finite electron-phonon coupling
$\lambda_{\rm MIT}$ where $Z \rightarrow 0$.
The metal-insulator transition is
accompanied by a softening of the phonon frequency $\Omega_0 \rightarrow 0$ for
a value $\lambda_{\rm ph}$ of the electron-phonon very close to
$\lambda_{\rm MIT}$, followed by a sudden increase for $\lambda >
\lambda_{\rm ph}$.\cite{bulla,koller,Max1}

The approach to the bipolaronic metal-insulator transition is reflected also
in the two-particle quantities $\chi$ and $\rho$.
In particular, the increase in the number 
of singlet local electron pairs close to the transition leads to a suppression
of the local spin susceptibility and
to an enhancement of the local charge susceptibility,
so that $\chi$ has a sharp drop whereas $\rho$ nearly diverges
at $\lambda_{\rm ph}$.

In this context the Hubbard repulsion
gives rise to
a further reduction of metallic properties at small
$\lambda$ (decrease of $Z$) while, on the other hand,
it disfavors double occupancy leading to a shift
to larger $\lambda$ of the critical values of the el-ph coupling
and to an increase (decrease) of the spin (charge) susceptibility.
Finally, increasing the quantum lattice fluctuations ($\gamma=1.0$,
right panels)
leads to a further increase of the critical electron-phonon
couplings.
\begin{figure}[!t!]
\centerline{\psfig{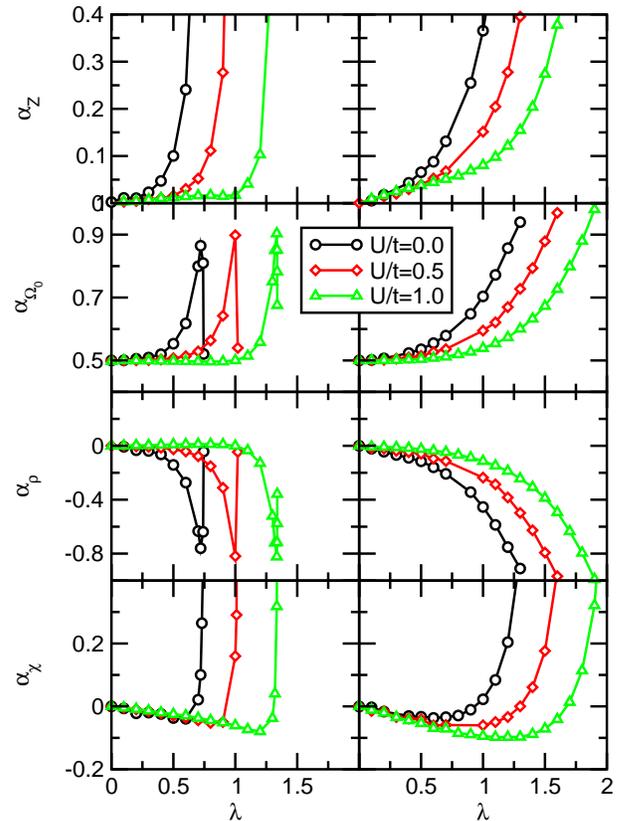}}
\caption{(color online)
Plot of the isotope coefficients for the same quantities
defined in Fig. \ref{f-Uweak}.} 
\label{f-isoUweak}
\end{figure}

In Fig. \ref{f-isoUweak} we show
the corresponding isotope coefficients
\begin{equation}
\label{def:alpha}
\alpha_X = - \Delta\log X / \Delta\log M
= (1/2) \Delta\log X /\Delta\log \gamma,
\end{equation}
where $X$ is $Z$, $\Omega_0$, $\chi$ and $\rho$.
We have chosen to compute the IEs using a finite value 
$\Delta \gamma/\gamma=0.15$.

The behavior of IEs approaching $\lambda_{\rm MIT}$ and 
$\lambda_{\rm ph}$ is depicted in Fig. \ref{f-isoUweak}.
The most striking effect is the near {\it divergence} of the isotope
coefficients as the system approaches the pairing metal-insulator transition.
As a matter of fact, while $\alpha_{Z}$ has a true divergence
as $\lambda \rightarrow \lambda_{\rm MIT}$, for the isotope effects
on the other quantities the divergence is cut-off by the not complete
suppression of charge fluctuations. From the practical point of view
however both the MIT and bipolaronic instabilities are accompanied
by a huge increase of the magnitude of the anomalous isotope effects.

The key point to understand this issue
is that an isotopic shift of $\omega_0$ is reflected in a finite
shift to higher $\lambda$ of both the bipolaronic
MIT ($\lambda_{\rm MIT}$) {\it and} the phonon softening ($\lambda_{\rm ph}$).
It obviously descends from  Eq. (\ref{def:alpha}) that
the divergence of a quantity is accompanied by a
divergence to $-\infty$ of the corresponding isotope coefficient
$\alpha$, while a positively 
divergent isotope coefficient is associated to a vanishing observable.
According this view we can understand the behaviors of Fig.  \ref{f-isoUweak}.
Namely we find that  $\alpha_{Z} \rightarrow \infty$
as $Z \rightarrow 0$ at $\lambda_{\rm MIT}$ and
$\alpha_{\chi}$ nearly diverges as $\chi$ rapidly drops
at $\lambda_{\rm MIT}$.
In similar way the near vanishing of
$\Omega_0$ and the near divergence of $\rho$, at $\lambda_{\rm ph}$
are reflected in $\alpha_{\Omega_0} \rightarrow +\infty$ and $\alpha_{\rho}
\rightarrow -\infty$. The following drop of $\alpha_{\Omega_0}$ and the jump of
$\alpha_{\rho}$ for $\lambda > \lambda_{\rm ph}$ reflect the sharp behavior
shown in Fig.\ref{f-Uweak} of the corresponding quantities.

Increasing $U$ leads to an increase of $\lambda_{\rm MIT}$, while increasing 
$\omega_0$ determines an overall smoothing of these features.
An interesting point to be noted here is the non monotonic behavior
of the IE on the spin susceptibility $\chi$, which
shows a negative sign at small $\lambda$ before approaching
a near divergence towards $+\infty$
close to the metal-insulator transition.
On the contrary $\alpha_{\rho}$ remains negative up to $\lambda_{\rm ph}$.
The negative sign of $\alpha_\chi$ and $\alpha_{\rho}$ 
for small $\lambda$ can be
understood in a perturbative framework. 
The electron-phonon interaction introduces
a positive Stoner-like contribution to the
effective spin susceptibility:
\begin{equation}
\chi = \frac{\chi_0}{1-I_{\rm el-el}+I^{\chi}_{\rm el-ph}},
\end{equation}
where $I_{\rm el-el} \propto U$ arises from the 
electron-electron repulsion
and the electron-phonon term $I^{\chi}_{\rm el-ph} \propto \gamma \lambda$
takes into account the explicitly isotope dependence
due to the finite bandwidth and vertex corrections arising
from the electron-phonon interaction.\cite{chi,cgp-chi} 
In the same RPA fashion the charge susceptibility reads
\begin{equation}
\rho = \frac{\rho_0}{1+I_{\rm el-el}+I^{\rho}_{\rm el-ph}},
\end{equation}
where  $I^{\rho}_{\rm el-ph} = -\lambda + I^{\chi}_{\rm el-ph}$. 
Thus at a perturbative level the only source of IE in $\rho$ and 
$\chi$  comes from $I^{\chi}_{\rm el-ph}$, which can be
shown to be the same as in both quantities 
leading to the {\it same} negative IE.

\begin{figure}[!t!]
\centerline{\psfig{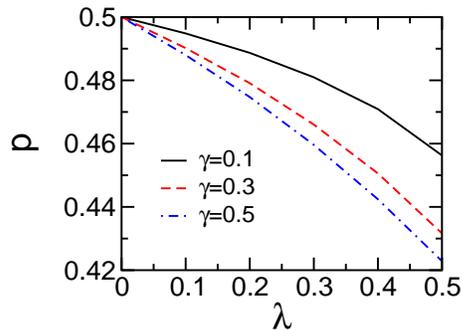}}
\caption{(color online) Behavior of single occupation number
$p=|\langle \uparrow |\phi_0 \rangle|^2 +
|\langle \downarrow |\phi_0 \rangle|^2$
as function of $\lambda$
in the weak el-ph coupling regime ($\lambda\simeq 0.5$)
for different adiabatic ratio $\gamma$ and for $U=0$.}
\label{f-p}
\end{figure}

The sign of the isotope coefficient on the spin susceptibility
reflects thus two different physical regimes where lattice fluctuations
triggered by $\omega_0$ play a different role. Namely,
in the perturbative small $\lambda$ regime
lattice fluctuations disfavor single spin occupancy
(Fig. \ref{f-p}),
leading to a decrease of $\chi$ and to a negative isotope coefficient.
On the other hand the physics in the $\lambda \lsim \lambda_{\rm MIT}$
region
is dominated by the proximity to the bipolaronic MIT. In this context
lattice fluctuations relax the phonon-induced electron trapping
favoring the unpaired spin singlet states.
Along this line
the sign of the isotope coefficient on $\chi$ can be thus used
as a useful trademark to characterized the different physical regimes.
The ability of the IE to characterize the physical
properties of different regimes will be even more remarkable
in the strongly correlated regime.

\section{Strongly correlated regime}
\label{s-strong}

In the previous section we have seen that for small $U$
the physical properties of the system
are qualitatively similar to those of the
pure Holstein model, where turning on a small
Coulomb repulsion leads essentially only to
an increase of the critical value of $\lambda$ for bipolaron pairing.
The situation is significantly changed in the strongly correlated
regime in which the Hubbard repulsion exceeds the electron-phonon coupling.
In this case a more reasonable starting point is the physics of 
the pure Hubbard model whose metallic region,
close to the Mott transition, can be described in terms of a narrow
resonance at Fermi level accompanied by the two high-energy Hubbard subbands.
In this framework the electron-phonon interaction
has two opposite effects: on one hand it provides a further
source of scattering which would decrease the quasi-particle scattering
weight $Z$; on the other hand it
acts as a effective attraction between the electrons, competing with the
Hubbard repulsion.\cite{sgiovanni}
In the strongly correlated regime, when the reduction
of the effective repulsion prevails over the
electron-phonon renormalization of the quasi-particle properties,
increasing $\lambda$ leads to an enhancement of $Z$.\cite{koller,sgiovanni}

Another interesting feature of the Hubbard-Holstein model
pointed out in Ref. \onlinecite{koller} in the strongly correlated regime
is the change of the order of
the bipolaronic MIT at $\gamma=0.1$.
The bipolaronic transition, which is continuous (second order)
for $U/t \lsim 1.5$, becomes indeed discontinuous (first order)
for $U \bsim 1.5$. In this latter regime
all the observable
have thus a jump at $\lambda_{\rm MIT}$.
The evolution from a second order to a first order transition
has been discussed in terms of a
Landau-Ginzburg theory where the first/second order transition is driven
by the local lattice configuration.\cite{jeon}
From a numerical point of view, however, it is hard to distinguish,
on the basis of observables like $Z$ or $\Omega_0$
a first order transition from a sharp
second order one. We show that the analysis of the phonon IE
is a powerful tool to distinguish in a very clear way
the first order or the second order character of the bipolaronic MIT.
This is even more interesting since,
as we are going to show,
the first order transition is not related to a phonon
mechanism while it is probably of a electronic origin.

\begin{figure}[!t!]
\centerline{\psfig{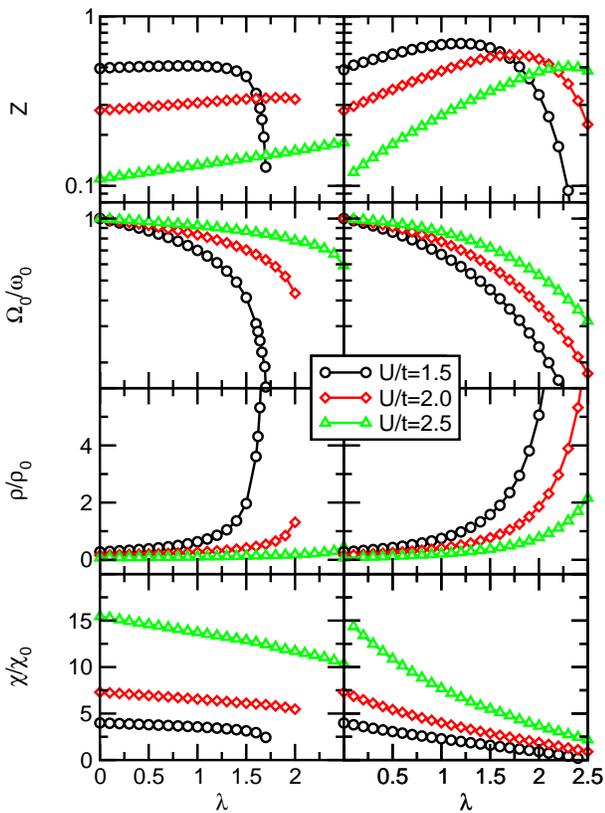}}
\caption{(color online) Behavior of the quasi-particle
spectral weight $Z$, of the
renormalized phonon frequency $\Omega_0$ and of the
charge and spin susceptibilities, $\rho$ and $\chi$,
as functions of
the electron-phonon coupling $\lambda$ for different intermediate-strong
values
of the Hubbard repulsion $U$. In the left panel
the curves are plotted for the adiabatic ratio
$\gamma=0.1$ while in the right panel the adiabatic ratio is
$\gamma=1.0$.}
\label{f-Ustrong}
\end{figure}

In Fig. \ref{f-Ustrong} we show $Z$, $\Omega_0$, $\chi$ and $\rho$
as function of $\lambda$
in the regime of strong correlation for $\gamma=0.1$ and for
$\gamma = 1.0$.
As mentioned above, in this strongly correlated regime,
increasing $\lambda$, as long as the system
does not reach a bipolaronic regime, results
in a reduction of the effective Hubbard repulsion and in an
increase of $Z$.
It is however interesting to note 
that the behavior of the other quantities, namely
$\Omega_0$, $\chi$ and $\rho$, as function of $\lambda$ and $U$ does not
change
qualitatively with respect to the weakly correlated case.
Concerning the character of the bipolaronic MIT,
we find that the
transition is always of the second order for any $U$ at
$\gamma = 1.0$.
On the other hand, due to the lack of precursor
effects in the metallic phase,
it is hard to predict the character of the transition
at $\gamma=0.1$.

\begin{figure}[!t!]
\centerline{\psfig{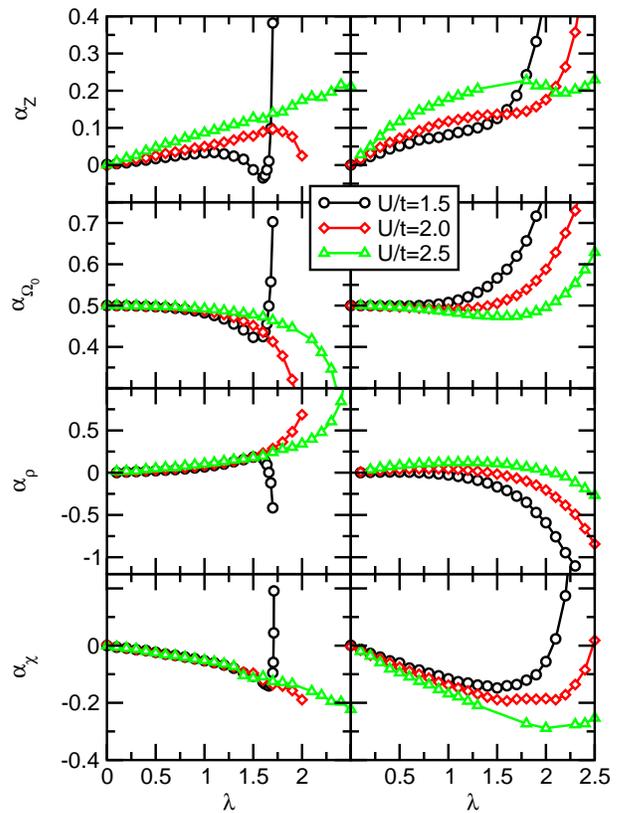}}
\caption{(color online)
Plot of the isotope coefficients for the same quantities
reported in Fig. \ref{f-Ustrong}.}
\label{f-isoUstrong}
\end{figure}

The corresponding IEs on the quantities
$Z$, $\Omega_0$, $\chi$ and $\rho$ is reported in
Fig. \ref{f-isoUstrong}.
We note a qualitatively different behavior of the isotope coefficients
in the strongly correlated regime at $\gamma=0.1$ as compared
with the weak $U$ regime and (partially) with the case $\gamma=1.0$.
Let us discuss for the moment the highly correlated cases $U/t=2.0,2.5$.
Most striking is the behavior of the phonon and of the charge
susceptibility
isotope coefficient which shows
a trend diametrically opposite to the weak-$U$ case,
namely a marked {\em reduction} (increase) of $\alpha_{\Omega_0}$
($\alpha_{\rho}$)
to be compared with the divergences
$\alpha_{\Omega_0}\rightarrow  \infty$,
$\alpha_{\rho}\rightarrow  -\infty$ for small $U$.
Among the other anomalous features we note
a typical downturn of $\alpha_{Z}$
as function of $\lambda$, which is less marked as $U$ increases.
On the contrary, the upturn of the isotope coefficient on the spin
susceptibility which is evident at small $U$ is here absent.
Quite interesting is also the $U/t=1.5$ case which
is very close to the value where the bipolaron MIT changes
from the second order to the first order.
In this case the IEs show
similar trends ($\alpha_{\Omega_0}<0.5$, $\alpha_{\rho}>0$, downturn of
$\alpha_{Z}$) as for the first order cases
in a large range of $\lambda$,  with a final upturn
(downturn for $\alpha_{\rho}$)
towards the characteristic divergences of the second order.
This trend is clearly visible also at $\gamma=1.0$ (although the
bipolaronic transition is always of the second order) pointing out
an partial underlying tendency towards a first order instability.
The larger amount of the lattice fluctuations in this case
as compared with $\gamma=0.1$ prevents however the occurrence of the
first order instability and enforces the second order character.

\begin{figure}[!t!]
\centerline{\psfig{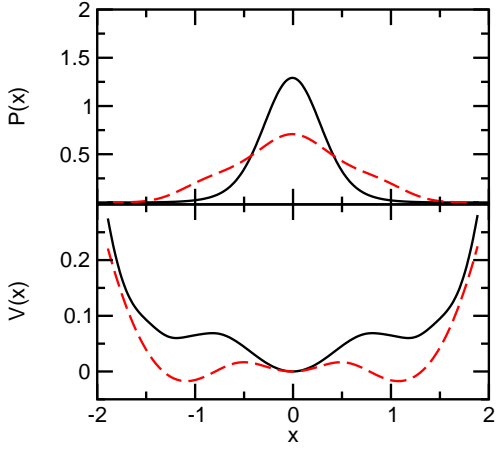}}
\caption{(color online) PDF lattice function $P(x)$ and lattice
potential $V(x)$.
Solid black line: $U/t=2.0$, $\lambda=2.0$, dashed red line:
$U/t=1.5$, $\lambda=1.7$.}
\label{f-pxvx}
\end{figure}

As we are going to show,  the physical origin of these
anomalous behaviors can be related to different phonon regimes,
although they do not rule directly the change between
the first and second order character
of the transition.
In order to clarify in more details this issue we evaluate the
local lattice probability distribution function (PDF)
$P(x)$ which provides information about
the lattice wavefunction of the ground state.
In Fig. \ref{f-pxvx} (upper panel) we show the PDF for two representative
cases characterized by significant lattice distortions
induced by the electron-phonon interaction,
resulting in a non-gaussian $P(x)$. In the strictly adiabatic case,
polaronic effects for instance are reflected by a bimodal shape of $P(x)$,
where
the two maxima are associated to different electron occupancies (e.g.,
in spinless systems $n=0$ and $n=1$, in spinful systems $n=0$
and $n=2$).
The broadness of these structures induced by
quantum lattice fluctuations however can make hard
to resolve the effective extent of the polaronic lattice distortions.
To overcome this problem
we extract an ``effective''
lattice potential $V(x)$ by solving the inverse Schr\"odinger problem
\begin{equation}
\left[-\frac{1}{2M}\nabla^2+V(x)\right]\bar{\psi}(x)=\omega_0\bar{\psi}(x),
\end{equation}
where $\bar{\psi}(x)=\sqrt{P(x)}$ and $M$ is the atomic mass.
Fig. \ref{f-pxvx} shows the lattice distribution function
$P(x)$ and the corresponding lattice potential $V(x)$
for two representative cases where the multi-valley structure
of $V(x)$ is not reflected in the shape of $P(x)$.
We stress that the above defined $V(x)$
reproduces the physical lattice potential only in the
strict adiabatic limit $\gamma \rightarrow 0$.
It can be considered however to be still a valid approximation
at finite and small $\gamma$ providing thus useful
insights for the true (unknown) lattice potential.

\begin{figure}[!t!]
\centerline{\psfig{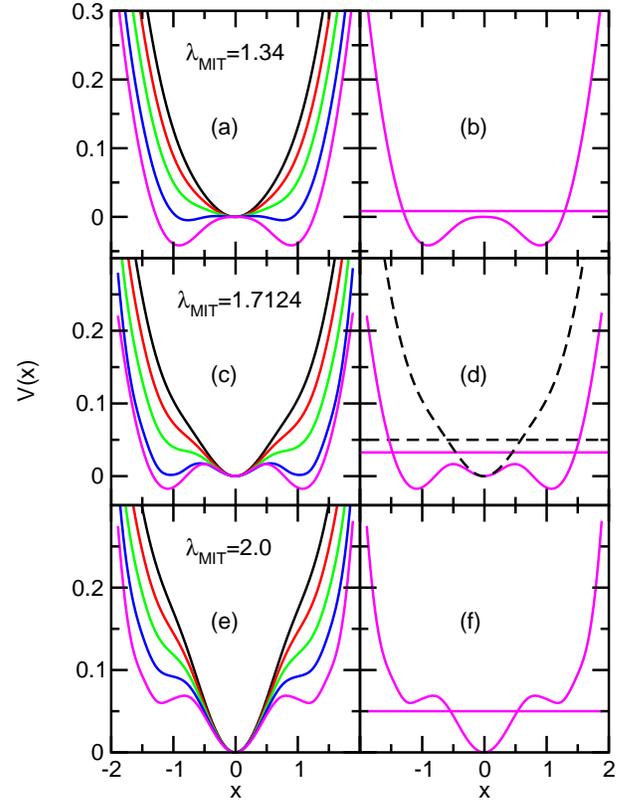}}
\caption{(color online) Lattice potential $V(x)$ in different physical regimes.
Upper panels: $U/t=1.0$, $\lambda=1.0, 1.1, 1.2, 1.3, 1.34$;
(right: $\lambda=1.34$).
Middle panels: $U/t=1.5$, $\lambda=1.4, 1.5, 1.6, 1.7, 1.71246$;
(right: $\lambda=1.4,,1.71246$).
Lower panels: $U/t=2.0$, $\lambda=1.6, 1.7, 1.8, 1.9, 2.0$;
(right: $\lambda=2,0$).
The horizontal lines in panels b,d,f marks the zero point
lattice energy for the corresponding lattice potential.
}
\label{f-psvx}
\end{figure}

In Fig. \ref{f-psvx} we report the evolution of the lattice potential
$V(x)$ as function of $\lambda$
for three representative cases of the Hubbard repulsion,
$U/t=1.0, 1.5, 2.0$, and for $\gamma=0.1$.
All the shown $\lambda$'s belong to the metallic phase. 
The largest value of $\lambda$, in each panel 
of the left side of Fig. \ref{f-psvx}, 
corresponds to the MIT.
 
The first case (panel a)
shows, in agreement with a second-order transition scenario,
a continuous flattening of the lattice potential as $\lambda$ increases.
This picture shows in the clearest way the lattice instability
at $\lambda_{\rm ph} \simeq 1.3$, where the quadratic term of
$V(x) \simeq a_2 x^2 + a_4 x^4 + \ldots$ vanishes
and the renormalized phonon frequency $\Omega_0 \rightarrow 0$.
For $\lambda_{\rm ph}<\lambda<\lambda_{\rm MIT}$ the systems is still metallic
and a double-well structure of the lattice potential is developing.
This corresponds to the phonon hardening back shown
in Fig. \ref{f-Uweak}.
The electronic metal-insulator transition occurs for
$\lambda_{\rm MIT} > \lambda_{\rm ph}$
when the double-well potential is already established
and the MIT is caused by the vanishing of coherent tunneling between
the two minima.
A crucial role is played by the zero point phonon energy
$\omega_0/2$ (horizontal  lines in panel b)
which favors tunneling processes.
In particular, in the latter situation where the double-well is
already formed, an increase of
$\omega_0\rightarrow \omega_0 + \Delta \omega_0$,
due to an isotope shift,
further enhances the metallic character, enhancing $Z$ and
resulting in a positive isotope coefficient on $Z$.
At the same time the system, increasing
$\omega_0$,
will probe a less shallow potential (panel b) resulting
in a stronger hardening of $\Omega_0$ than the harmonic one and
leading to $\alpha_{\Omega_0}>0.5$.

A more complex situation is found for $U/t=1.5$.
Here, more than a softening of the quadratic term $a_2 x^2$,
increasing $\lambda$ the lattice potential $V(x)$ presents
two flat regions at finite $x$ which evolve first in
two relative minima and then in two absolute minima
lower than the central one (panel c).
In this framework we distinguish two regimes.
The first one is when the zero point energy is roughly lower than
the flat regions (or minima) of the potential ($\lambda \lsim 1.6$);
in this situation increasing $\omega_0$ makes the lattice wave-function
more distorted, the metallic character is reduced and the phonon energy
is softened with respect to the harmonic one.
This phenomenology is thus reflected in the downturn of
$\alpha_{Z}$ and of
$\alpha_{\Omega_0}<0.5$.
A second regime is reached for $\lambda$ closer to the MIT
($\lambda \bsim 1.6$): here the minima are lower than the zero point
energy,
a phenomenology similar to the double-well potential is restored and
$\alpha_{Z}> 0$, $\alpha_{\Omega_0}>0.5$.

The results for $U/t=2.0$ (panels e,f) are essentially similar to those for
$U/t=1.5$ for small $\lambda$. Here however the first-order metal-insulator
transition occurs well before the relative minima at finite $x$
reach the range of the zero point energy.
As a consequence the previous considerations apply, namely
$\alpha_{Z}$ shows an downturn (but it still remains finite)
and $\alpha_{\Omega_0}<0.5$.

The evolution of the lattice potential as function
of $U$ and $\lambda$ sheds interesting light on
the origin of the first-order/second-order character of the MIT.
It has been indeed claimed that both the second-order at small $U$
and the first-order transition at large $U$ could be related to a
phonon-driven mechanism.\cite{jeon}
In a phonon-driven Ginzburg-Landau theory the character of the transition
is thus determined by the dependence of the free energy
on the lattice coordinate,
$F(x) \approx a_2 x^2/2 + a_4 x^4/4 + a_6 x^6/6$, which in the
adiabatic limit reduces to the lattice potential $V(x)$.
The second-order transition at small $U$
is thus associated with the development of
a double-well potential which occurs
for $a_2 < 0$ and $a_4, a_6 > 0$, whereas
a first order one is expected to correspond to a three-well structure
with absolute minima at $x_0 \neq 0$, condition realized for
$a_2, a_6>0$, $a_4 < -\sqrt{4a_2 a_6}$.
Our results shown in Fig. \ref{f-psvx}
confute however this picture.

At $U \lsim 1.5$
the second order MIT transition occurs {\it after} the development of the
double well in $V(x)$. 
MIT here  is indeed a by-product of the lattice distortions as was shown by
adiabatic analysis of Ref. \onlinecite{millis2} and is due 
to
the vanishing of the coherent tunneling
processes between the minima of two double-well lattice potential.

For $U \bsim 1.5$  instead the developing of the three-well shape
of the lattice potential
is not sufficient to explain the
first-order character since the MIT occurs far {\it before} the
local minima with finite lattice
distortions at $x \neq 0$ become absolute minima.
Moreover the energy of the local minima at the transition
become higher and higher
in energy as $U$ increases, precisely where the first-order character
becomes more marked.
These considerations suggest thus that the
origin of the first-order metal-insulator transition
in the strongly correlated regime is likely ascribed
to an electronic origin more than some phonon-driven
instability. 
An intrinsic difference between the weakly and the strongly  correlated
metallic phase is the presence of a spin-singlet character in the metallic case
pointed out by the presence of the Hubbard bands coexisting with the Kondo
peak in the strongly correlated regime.  
If the metallic state has a pronounced spin-singlet character the transition to
a bipolaronic pair state (which is instead a charge pseudo-spin singlet) 
can be viewed as a level crossing between this two different electronic states
therefore leading to
a first order transition.

\section{$U$-dependence}
\label{U-dependence}

In the previous section we have analyzed in details
the behavior of the IEs on the physical quantities
as function of the electron-phonon coupling $\lambda$.
We have shown in particular that the IEs
are quite sensitive to the different
regimes of the electronic correlations, and we have distinguished
the weak from the highly correlated regime.
The aim of this section is to address this issue in more details
by studying the dependence of the IEs on the Hubbard repulsion $U$.
We shall focus on the IE on the quasi-particle spectral weight $Z$
and on the renormalized phonon frequency $\Omega_0$ which show the most
characteristic features.
We consider two values of $\lambda$,
representative of two different physical regimes:
$\lambda=0.7$, for which the system is far from the bipolaronic
phase
even in the absence of any Hubbard repulsion;
and $\lambda=1.5$ for which value instead
the ground state is a bipolaronic insulator
at $U=0$. In this case one needs a finite value of
the Hubbard repulsion ($U \lsim 1.25$) to contrast the el-ph driven
attraction
and to restore the metallic character. Both the cases
of course undergo a Mott metal-insulator transition as $U$ approaches
some critical value $U_c$.

\begin{figure}[t]
\centerline{\psfig{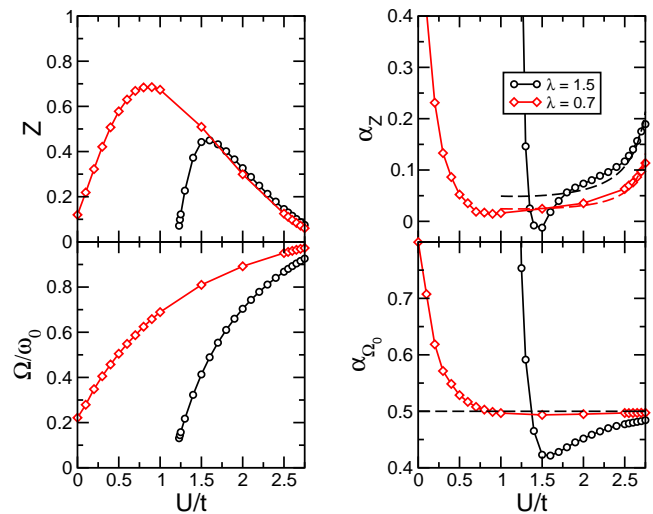}}
\caption{(color online) Left panel: quasi-particle spectral weight $Z$ and
renormalized phonon frequency $\Omega_0$
as functions of Hubbard repulsion $U$ for two different $\lambda=0.7$
(empty diamonds) and $\lambda=1.5$ (empty circles).
Right panel: the corresponding IEs
$\alpha_{Z}$ and $\alpha_{\Omega_0}$.
All curves are plotted for $\omega_0/t=0.1$.
In the right-top panel are also shown the analytical behavior
predicted by Eq. (\ref{analytical}) (dashed lines) by using,
respectively,
$U_{c2}/t=2.87$ for $\lambda=0.7$ and $U_{c2}/t=2.90$ for $\lambda=1.5$.
}
\label{trend-U}
\end{figure}
In Fig. \ref{trend-U} we report the evolution of the quasi-particle
spectral weight $Z$
and of the renormalized phonon frequency $\Omega_0$
as functions of $U$ for the two chosen $\lambda$.
Let us discuss first the behavior of
$\Omega_0$, which
shows a monotonic behavior for $\lambda=0.7$. This trend is
representative for the generic case of weak electron-phonon coupling:
due to the suppression of the double/empty sites,
increasing $U$ leads to a reduction of the charge fluctuations and to
the consequent enhancement of the renormalized phonon frequency $\Omega_0$
which results less screened.
More intriguing is the analysis of $Z$
which, unlike the pure Hubbard model, shows a non-monotonic
behavior as function of $U$.
In the previous section we have discussed how, in the strongly correlated
regime close to the Mott MIT,
switching on the electron-phonon interaction leads to an increase
of $Z$. This trend has been explained as a result of the competition
between the attractive el-ph coupling
and the repulsive Hubbard term.\cite{koller,sgiovanni}
Fig. \ref{trend-U} shows that a similar mechanism applies also
close to  the bipolaron MIT where the phonon-driven attraction is the strongest
interaction, and the Hubbard repulsion partially counteracts it,
so that $Z$ increases with the repulsion for small $U$.
For larger $U$ however the repulsion overcomes the attraction,
the system approaches the Mott MIT and $Z$ recovers the standard 
behavior ($Z \rightarrow 0$).
These trends are also more marked for $\lambda=1.5$ where, as mentioned
above, the system is a (bipolaronic) insulator in the absence
of Hubbard repulsion, and a finite $U \bsim 1.2$ is required
to have a metallic state.
Most evident is of course the competition between $U$ and
the electron-phonon interaction, which becomes a sharp transition
from $Z=0$ for $U \lsim 1.2$ to a maximum value
$Z \simeq 0.5$ for $U \simeq 1.5$.

In the right panels of Fig. \ref{trend-U} we show the
corresponding IEs $\alpha_{Z}$,
$\alpha_{\Omega_0}$. Also here the behavior of the isotope
effect on the quasi-particle spectral weight is particularly
interesting. In section \ref{s-weak} we have shown that giant
IEs on the quasi-particle spectral weight $Z$ are expected as the system
approaches the bipolaronic metal-insulator transition. From an intuitive point
of view we can expect that the electronic correlation, driving
the systems far from the bipolaronic transition, reduces the
magnitude of the anomalous IE $\alpha_{Z}$. This
picture is confirmed in Fig. \ref{trend-U} where we see that
$\alpha_{Z}$ is strongly reduced as soon as
$U$ moves the system away from the bipolaronic MIT.
This trend is however quite soon followed by a further enhancement
of the anomalous IE on $Z$ which
becomes larger as the systems approaches the Mott transition,
revealing that the strongly correlated regime close to the Mott
metal-insulator transition is also highly sensitive to the phonon
dynamics. This counterintuitive result can be qualitatively
understood by following the analysis of Ref.
\onlinecite{sgiovanni} which showed that the strongly correlated
regime close to the Mott MIT {\em in the presence} of
electron-phonon interaction could be described in terms of a pure
Hubbard model with a properly rescaled Hubbard repulsion $U_{\rm
eff}=U-\eta\lambda t$, where $\eta=2\omega_0/(2\omega_0+U)$.
Since the quasi-particle spectral weight $Z$ is
known to scale linearly with $U_{c2}-U$ in DMFT for $U
\rightarrow U_{2c}$, where $U_{c2}$ is the critical value of the
Hubbard repulsion at which a Mott metal-insulator transition
occurs, we obtain thus
\begin{equation}
\alpha_{Z}=
\frac{\lambda \omega_0 t U}{(U+2\omega_0)^2(U_{c2}-U_{\rm eff})}.
\label{analytical}
\end{equation}
This expression shows that
the isotope effect on $Z$ diverges as $1/Z$
as $U_{\rm eff}\rightarrow U_{c2}$.\cite{sgiovanni,giorgionew}
The analytical behavior predicted by Eq. \ref{analytical},
is also shown in the right-top panel of  Fig. \ref{trend-U}
(dashed lines). The critical values $U_{c2}$ 
for each $\lambda$ have been chosen
in order to have the best fit with the DMFT numerical results:
namely
$U_{c2}/t=2.87$ for $\lambda=0.7$ and 
$U_{c2}/t=2.90$ for $\lambda=1.5$.

The crossover between the two different correlation regimes
is also pointed out by the analysis of the behavior
of the IEs on the
renormalized phonon frequency $\alpha_{\Omega_0}$.
Most evident is the case of $\lambda=1.5$
where in the absence of Hubbard repulsion the
ground state is a bipolaronic insulator.
Here the non-monotonic behavior of
$\alpha_{\Omega_0}$ reflects two different regimes:
$i$) a first sharp region $U \lsim 1.36$ where
$\alpha_{\Omega_0}\ge 0.5$.
In this regime
the system is characterized
by a double-well lattice potential with a zero point energy larger than
the barrier between the two minima
as shown in Fig. \ref{f-psvx}b.
By increasing $\omega_0$
the system will probe thus a less shallow potential resulting
in a stronger hardening of $\Omega_0$ than the harmonic one and
leading to $\alpha_{\Omega_0}>0.5$.
$ii$) for $U \bsim 1.36$, on the other hand, the shape
of the lattice potential $V(x)$ is more similar to
the panel (f) of Fig. \ref{f-psvx} where the zero point energy
is lower than the unstable minima. Increasing $\omega_0$
the system would thus probe the flat potential regions around
these two minima leading to an increase of the
renormalized phonon frequency smaller than the harmonic one
($\alpha_{\Omega_0}<0.5$).
These effects become smaller as the Hubbard repulsion $U$ increases since
the minima shift higher in energy and the IE
$\alpha_{\Omega_0}$ approaches asymptotically $0.5$
for $U \rightarrow \infty$.

A similar behavior occurs for $\lambda=0.7$ where however
the system in the absence of electronic correlation
is still metallic although close to the bipolaronic MIT.
In this case however increasing $U$ leads to
a reduction of the IE on $\Omega_0$ until,
for $U \bsim 0.85$, $\alpha_{\Omega_0}$ becomes smaller than $0.5$,
although in a less marked way than in the $\lambda=1.5$ case
because of the smaller electron-phonon coupling.
The evidence of this anomalous IE $\alpha_{\Omega_0}<0.5$
even for small $\lambda$ suggests thus that
this regime is not related to the closeness to the
bipolaronic metal-insulator transition.
It is worth to stress in addition that
the evidence of these two regimes of correlation
prompts out from the analysis of the IE  $\alpha_{\Omega_0}$
while the renormalized phonon frequency itself  $\Omega_0$
does not show any signature of this crossover.

\section{Conclusions}
\label{concl}

In this paper we have studied in details the isotope dependence of several
physical quantities: the quasi-particle spectral weight,
the renormalized phonon frequency,
and the static component of local  charge and spin susceptibilities in the
metallic regime of the Hubbard-Holstein model within DMFT approximation.
The work is carried out for the half-filled model
in the metallic phase without broken symmetries showing
anomalous behaviors approaching the pair
and the Mott metal-insulator transitions. We have
shown that the anomalies of isotope coefficients can be helpful in 
revealing different physical regimes. 
In particular, we have found that in the adiabatic regime $\gamma=0.1$
we can identify two different regimes according to the degree
of correlation.
For $U \lesssim 1.5$ the physical properties of the system are qualitatively 
similar to those of the pure Holstein model where the approach to the
bipolaronic metal-insulator transition obtained within DMFT 
is reflected in a near divergence of 
the IEs characteristic of bipolaronic binding.
In this regime we find a divergent $\alpha_{Z}$ and a strong increase of 
$\alpha_{\chi}$ close to the pair MIT while both $\alpha_{\Omega_0}$ and 
$\alpha_{\rho}$ have a near divergence at $\lambda_{\rm ph}$ 
($\alpha_{\Omega_0} \rightarrow +\infty$ and $\alpha_{\rho}
\rightarrow -\infty$).
The second regime establishes for $U \gtrsim 1.5$, where the bipolaronic MIT 
becomes of the first order.

The change in the order of the transition can be characterized in an accurate
way by means of our study of IEs.
In the highly correlated regime, the  isotope
effects are indeed not diverging even when the
first-order transition is approached. 
The lack of any divergence as well as an
analysis of the lattice potential suggest that the  first-order character
is more likely ascribed to an electronic mechanism than to a lattice-driven
instability. In addition we have shown that approaching the Mott-Hubbard MIT
also gives rise to drastic anomalous IEs.

\section*{Acknowledgments}

We acknowledge useful discussions with C. Castellani, L. Pietronero and  
G. Sangiovanni.
This work was partially funded by 
the MIUR project FIRB RBAU017S8R and PRIN projects
2003 and 2005.


\begin{thebibliography}{99}
\bibitem{notat}
In the infinite coordination limit we rescale
the hopping matrix element $t$ as $t \rightarrow t/2\sqrt{z}$,
so that the hopping amplitude $t$ determines the half-bandwidth.
\bibitem{note-infinte-d}
{Within DMFT the bosonic pairs that are formed in strong coupling
 have no dispersion. This leads to localization of such boson pairs, and to
the metal-insulator transition. 
The bosons can instead move in finite dimensions, eventually leading
to finite pair conductivity if the motion becomes coherent.}
\end{thebibliography}
\end{document}